# Satellite Images Analysis with Symbolic Time Series: A Case Study of the Algerian Zone

[1,2] Dalila Attaf, [1]Djamila Hamdadou
[1]Oran1 Ahmed Ben Bella University, LIO Department
[2]Spatial Techniques Centre
Oran, Algeria
dalila.attaf@gmail.com, dzhamdadoud@yahoo.fr

[3]Sidahmed Benabderrahmane, [2]Aicha Lafrid
[3]Paris 8 University, [2]Spatial Techniques Centre
[3]Paris, France, [2]Oran, Algeria
sidahmed.benabderrahmane@iut.univ-paris8.fr
lafridaicha@gmail.com

***Abstract*—Satellite Image Time Series (SITS) are an important source of information for studying land occupation and its evolution. Indeed, the very large volumes of digital data stored, usually are not ready to a direct analysis. In order to both reduce the dimensionality and information extraction, time series data mining generally gives rise to change of time series representation. In an objective of information intelligibility extracted from the representation change, we may use symbolic representations of time series. Many high level representations of time series have been proposed for data mining, including Fourier transforms, wavelets, piecewise polynomial models, etc. Many researchers have also considered symbolic representations of time series, noting that such representations would potentiality allow researchers to avail of the wealth of data structures and algorithms from the text processing and bioinformatics communities. We present in this work, one of the main symbolic representation methods "SAX"(Symbolic Aggregate Approximation) and we experience this method to symbolize and reduce the dimensionality of a Satellite Image Times Series acquired over a period of 5 years by characterizing the evolution of a vegetation index (NDVI).**

## I. Introduction

Satellite Images Time Series (SITS) are important information sources on the territory evolution. The study of these images allows to understand changes in specific zones but also to discover large-scale evolution patterns. However, discovering these phenomena imposes to respond to several challenges which are related to SITS characteristics and their constraints. Each pixel of a satellite image is described by several values and the evolution patterns are ported for very long a period which generates a very large volume of data which makes the information extraction complex and difficult. In order to both reduce the dimensionality of SITS and the information extraction, Time Series Data Mining generally gives rise to a change of time series representation. In this work, we present one of the main principal symbolic representation methods and we apply it on a SITS acquired over a period of 5 years. This representation will be used later in the change detection (such as vegetation evolution, buildings detection, etc.). We have used a vegetation index, NDVI (Normalized Difference Vegetation Index), which is related to the vegetation cover structure, particularly in the coverage rate of soil by vegetation, and its pigment content. Moreover, the NDVI vegetation index is the most widely used by the scientific community to study the vegetation. It is defined by NDVI = (PIR - R) / (NIR + R) PIR and R reflectance in the near infrared and red, respectively.

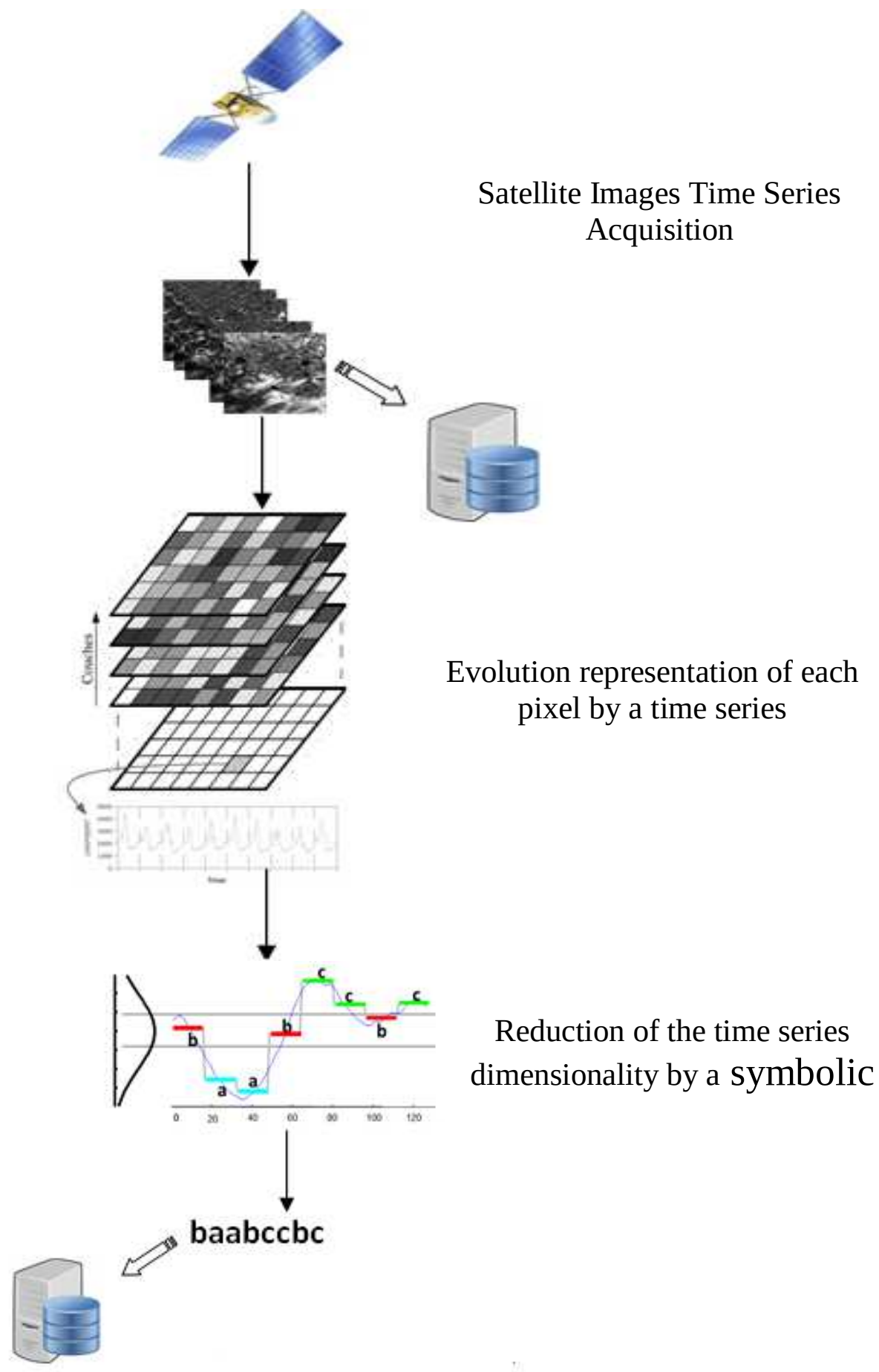

Fig. 1. Illustration of the studied problematic

## II. TIME SERIES DEFINITION

A time series is a collection of observations made sequentially in time [6].

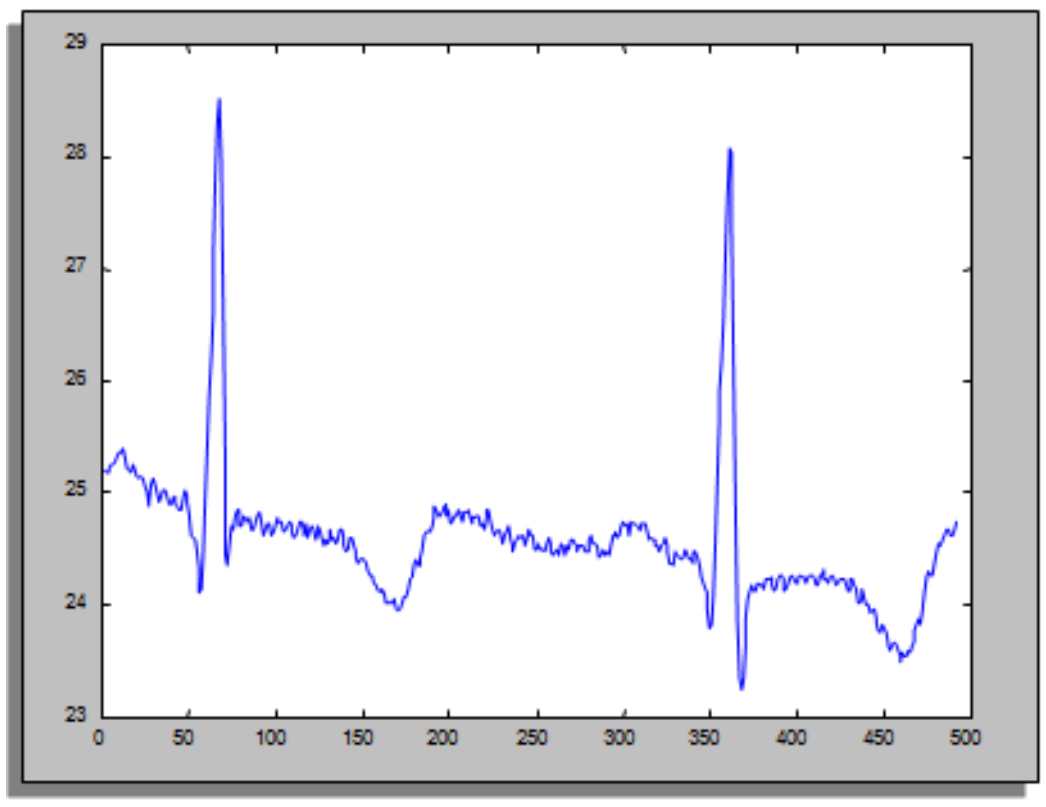

Fig. 2. Example of time series

Examples of time series include voice data, historical stock prices, sales histories, records of tests of an engine, the seismic data, flight records of aircraft, weather data, environmental data, satellite data, astrophysics data, etc [4].

## III. RELATED WORKS

Time series are data types that appear in many applications. Time series data mining includes many tasks such as classification, clustering, similarity search, motif discovery, anomaly detection, and others. Research in time series data mining has focused on two aspects; the first aspect is the dimensionality reduction techniques that can represent the time series efficiently and effectively at lower-dimensional spaces.

Different indexing structures are used to handle time series. Time series are high dimensional data, so even indexing structures can fail in handling these data because of what is known as the “dimensionality curse” phenomenon. One of the best solutions to deal with this phenomenon is to utilize a dimensionality reduction method to reduce the dimensionality of the time series, then to use a suitable indexing structure on the reduced space.

There have been different suggestions to represent time series in lower dimensional spaces. To mention a few; Discrete Fourier Transform (DFT) (Agrawal et al . 1993) and (Agrawal et al . 1995), Discrete Wavelet Transform (DWT) (Chan and Fu 1999), Singular Value Decomposition (SVD) (Korn et al . 1997), Adaptive Piecewise Constant Approximation (APCA) (Keogh et al . 2001), Piecewise Aggregate Approximation (PAA) (Keogh et al . 2000) and ( Yi and Faloutsos 2000), Piecewise Linear Approximation (PLA) (Morinaka et al . 2001), Chebyshev Polynomials (CP) (Cai and Ng 2004). [5]

Among the different representation methods, symbolic representation has several advantages, because it allows researchers to benefit from textretrieval algorithms and techniques that are widely used in the text mining and bioinformatics communities (Keogh et al . 2001).

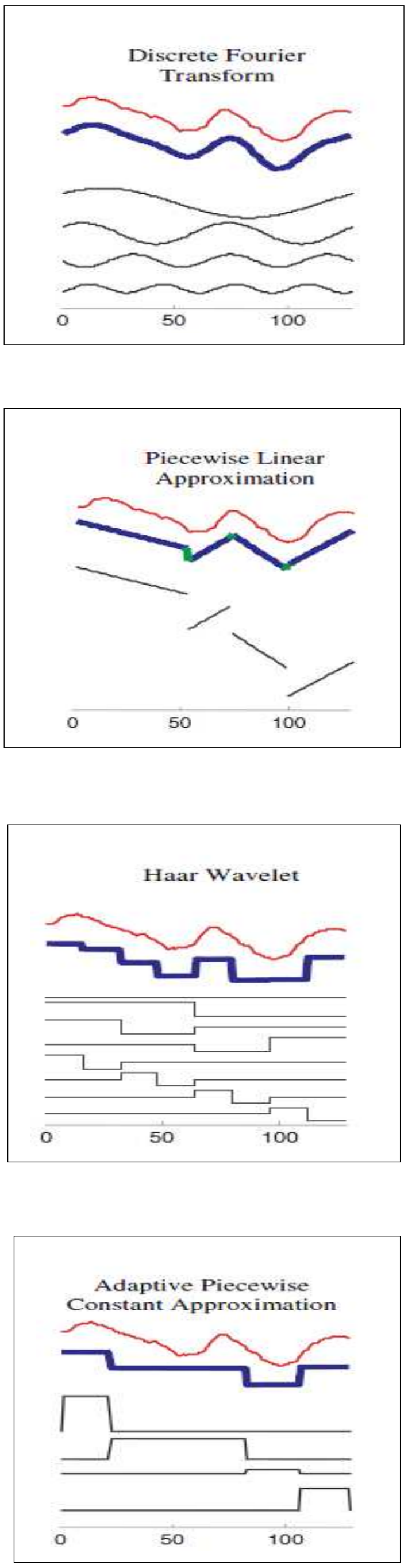

Fig. 3. The most common representations for time series data mining. Each can be visualized as an attempt to approximate the signal with a linear combination of basis functions

## IV. SYMBOLIC REPRESENTATION TIME SERIES

The purpose of changing the time series representation is often the dimensionality reduction. We seek to construct representation which preserves maximum information present in data without having any knowledge on what constitute this information.

We will present initially a generic framework of symbolic representation time series, before detailing the method which we have chosen for our experiments.

We have a given time series defined by ST defined by ST = $\{(d_i, v_i)\}_{i\in\{1\cdots N\}}$, with $d_i \in D$ et $v_i \in V$, with D the temporal definition domain and V is the digital space values of the time series.

We propose to define a symbolic representation of ST by:

- A division into P episodes.
- E = $\{e_j = [d_{jdebut}, d_{jfin}[\}_{\in j\{1\cdots P\}}$, and $(d_{jdebut}, d_{jfin}) \in D^2$ and $d_{jdebut} < d_{jfin}$
- An alphabet of symbols K $\Lambda = \{s_m\}_{m\in\{1\cdots K\}}$,
- Symbolic representation called SR : E $\to\Lambda$, ej $\mapsto$ RS(ej) is then the application associating each episode of E a symbol of .

A symbolic representation will be relevant if it best satisfies the following contradictory imperatives:

- Maximum concision representation: with cardinality of E, minimal $\Lambda$ and symbols as simple as possible.
- Maximum fidelity: it allows a reconstruction as near as possible to the original series [2]

## V. TIME SERIES SYMBOLIC REPRESENTATION BY SAX (SYMBOLIC AGGREGATE APPROXIMATION)

In order to reduce the dimensionality, it is important to define time units for grouping the time series points. We generally define episodes like intervals of time definition domain of time series to be represented. Generally, because of the minimal cost of acquisition and storage data, time series are stored in databases in the most detailed possible form, regardless of the time scale at which develop behaviors to identify. We can then group the points to episodes without losing essential information. This is the principle of symbolic representation SAX presented by Lin et al. (2003).

SAX is a time series symbolic representation univariate centered reduced:

- The time domain is divided into episodes of equal size.
- The equivalence classes of the time series values are set in advance according to the number of symbols to be used, so as to obtain a cutting of the same effective classes provided that the distribution centered and reduced values is normal [2].
- It allows a time series of arbitrary length n to be reduced to a string of arbitrary length w (w<<n) [6].

The SAX method is based on two main steps

### A. *Step1: Dimensionality Reduction via PAA (Piecewise Aggregate Approximation)*

A time series C = c1… cn of length n can be represented in a w-dimensional space by a vector $\bar{C} = \bar{c}_1 \ldots \bar{c}_w$.

The $i^{th}$ element of $\bar{C}$ is calculated by the following equation:

$$\bar{c}_i = \frac{w}{n} \sum_{j=\frac{n}{w}(i-1)+1}^{\frac{n}{w}i} c_j \qquad (1)$$

Simply stated, to reduce the time series from n dimensions to w dimensions, the data is divided into w equal sized “frames.” The mean value of the data falling within a frame is calculated and a vector of these values becomes the data-reduced representation. The representation can be visualized as an attempt to approximate the original time series with a linear combination of box basis functions as shown in Figure 4.

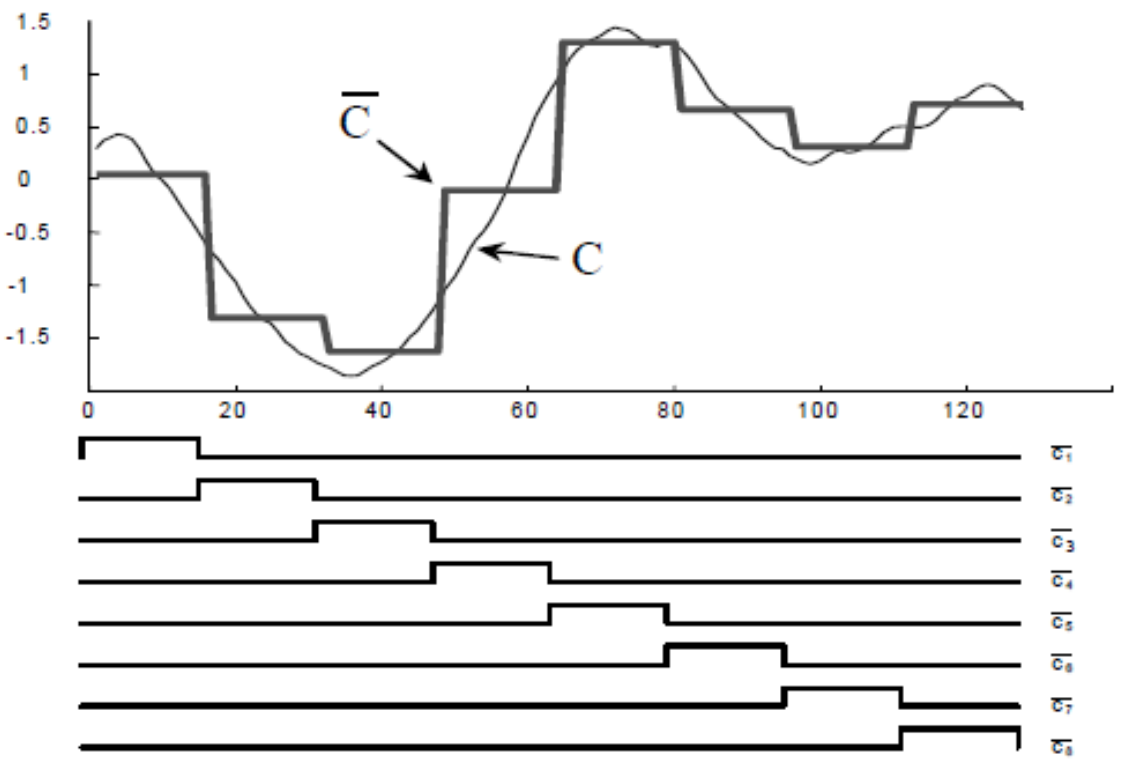

Fig. 4. The PAA representation can be visualized as an attempt to model a time series with a linear combination of box basis functions. In this case, a sequence of length 128 is reduced to 8 dimensions [8].

The PAA dimensionality reduction is intuitive and simple, yet has been shown to rival more sophisticated dimensionality reduction techniques like Fourier transforms and wavelets.

### *B. Step2 : Discretization*

SAX makes the assumption that time series values follow a Gaussian distribution. The quantization step makes use of (N-1) breakpoints (Gaussian quantiles) that divide the area under the Gaussian distribution into N equiprobable areas. These breakpoints can be found in lookup tables. Hence, the average values computed for each segment of the time series (step 2 above) are then quantized according to the breakpoints of the Gaussian distribution. Figure 5 shows an example of the SAX representation of a time series.

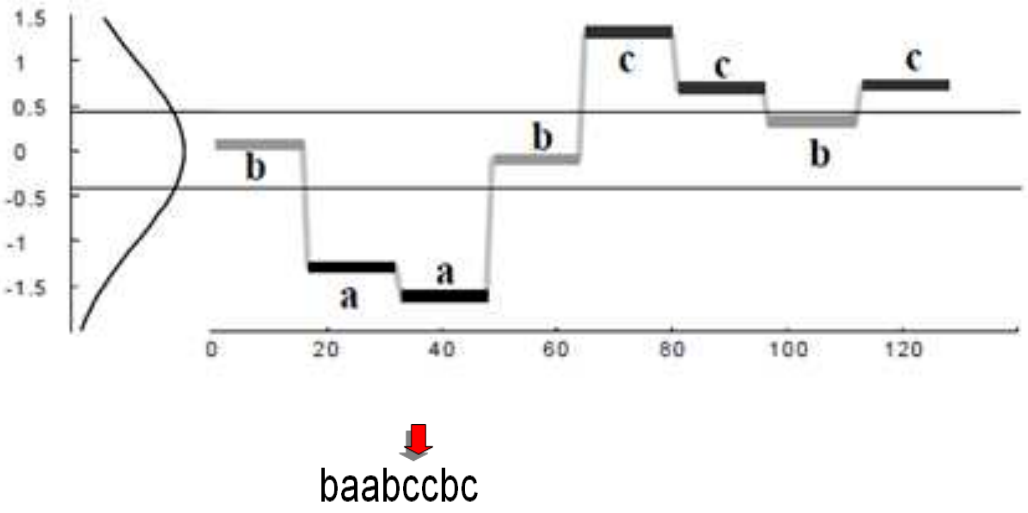

Fig. 5. Time series is discretized by first obtaining a PAA approximation and then using predetermined breakpoints to map the PAA coefficients into SAX symbols. In the example above, with n =128, w = 8 and a = 3, the time series is mapped to the word baabccb[8]

The figure 6 shows in details the SAX method working:

(1) The original time series

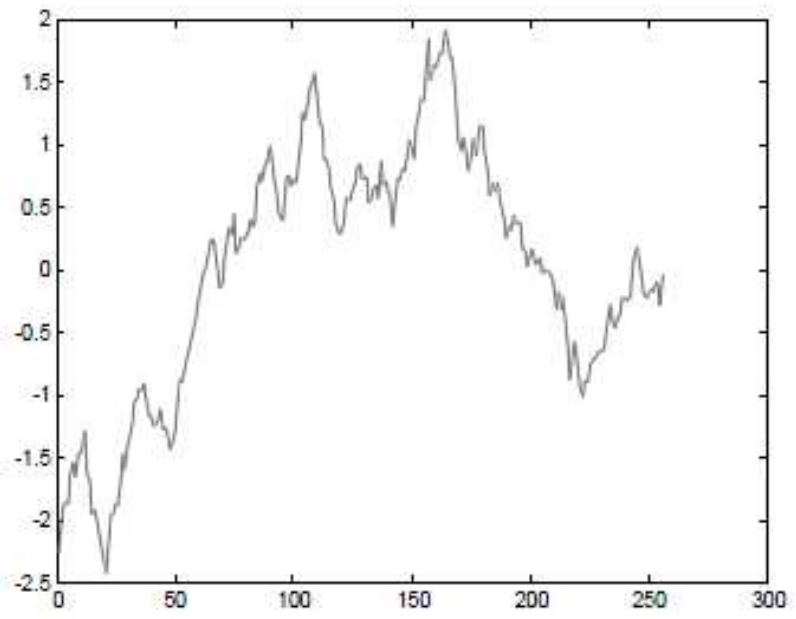

(2) Converting the time series to PAA

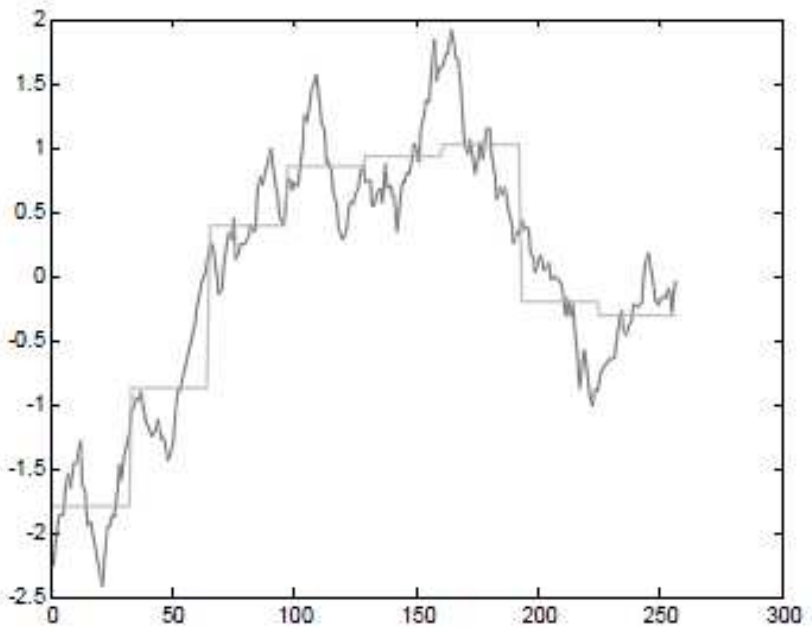

(3) Choosing the breakpoints

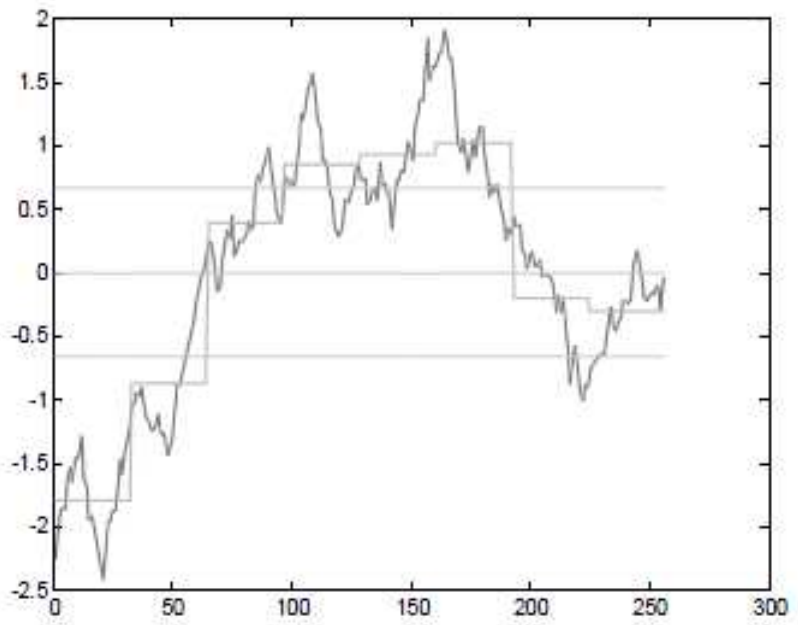

(4) Discretizing PAA

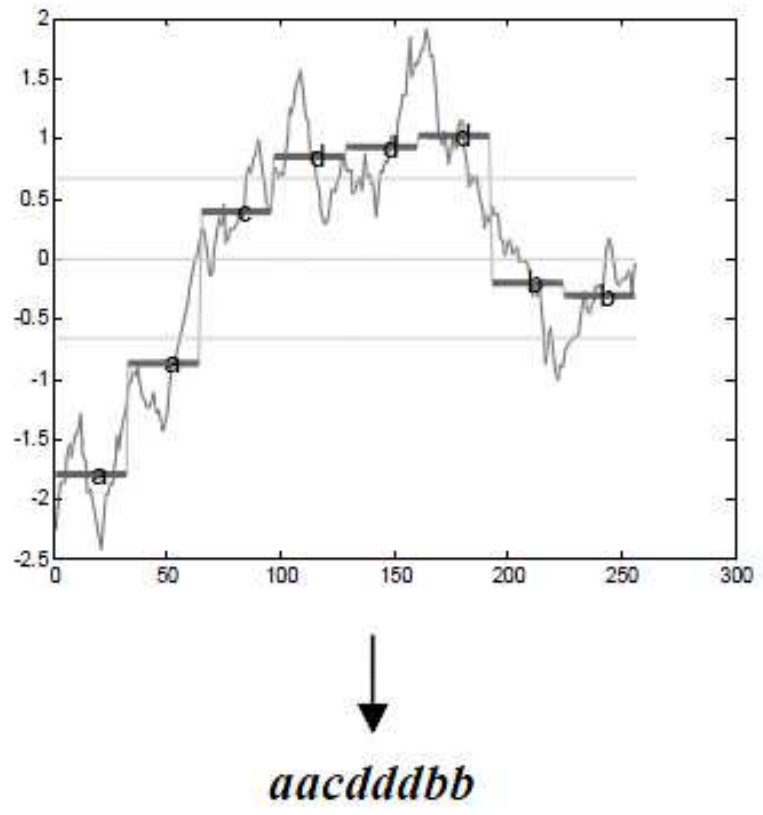

Fig. 6. The different functioning steps of applying SAX [4]

SAX method's advantages are:

- The construction of representations is extremely efficient in computation time (O (N)) for a time series representation of N points.
- All representations (based on the same number of symbols and episodes of the same size) are trivially commensurable. [2]
- Simplicity
- Need any a priori knowledge on data
- For two sets S1, S2 and their SAX representation Š1, Š2, we have dist(Š1, Š2) ≤ dist(S1, S2)

- Well adapted and efficient for classification, clustering and indexing. [3]

However, this performance suffers of disadvantages which are intrinsically linked:

- The modeling error for a given dimensionality reduction is not minimal since the model is not locally adapted to the data.
- Equivalence classes which symbols are associated, are not necessarily relevant because they are not adapted to the data. [2]
- Few semantics symbols. [3]

## VI. Implementation

Our approach proposed seeks to build Symbolic Satellite Images Time Series (SITS) using SAX method, in order to reduce the dimensionality of the original SITS and make them easier to interpret. The set of images constitute a SITS, and each pixel is associated to a time series. As shown in Figure 7 (Above, vision of layers images, and below, transversal vision of a pixel in the form of time series).

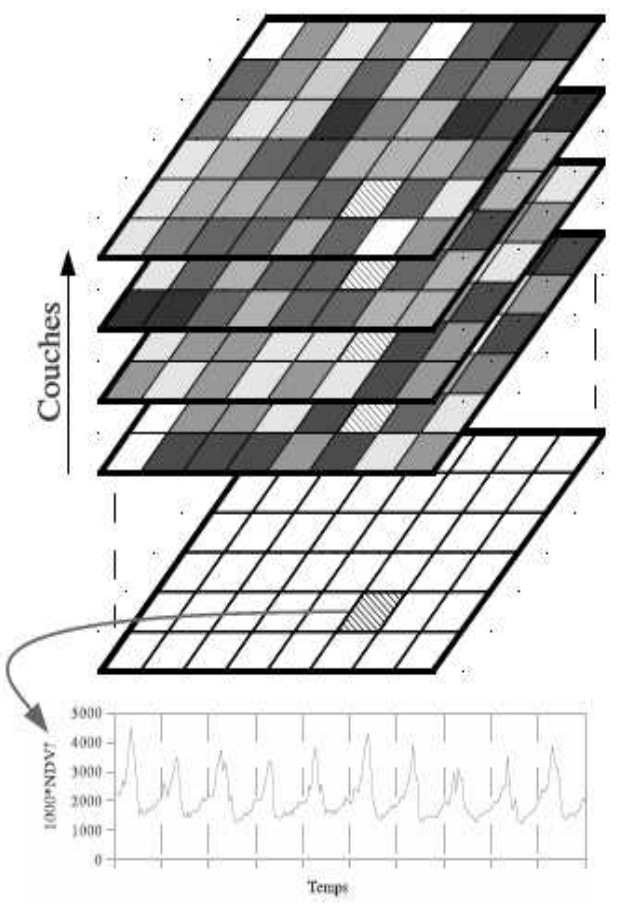

Fig. 7. Satellite Images Time Series (SITS) [1]

### *A. Data*

Satellite images used in this implementation are NDVI images of the satellite Terra/MODIS (Moderate Resolution Imaging Spectroradiometer) based on the evolution of vegetation indices (NDVI). These images are available in five years with important acquisition frequency (one image every sixteen days) and form so satellite image time series (SITS), which extends from the year 2000 to 2005 covering the same scene representing a part of an east Algerian city "Batna". Each image has 521X455 pixel's size with a spatial resolution of 250mX250m. In total, our database contains 135 images.

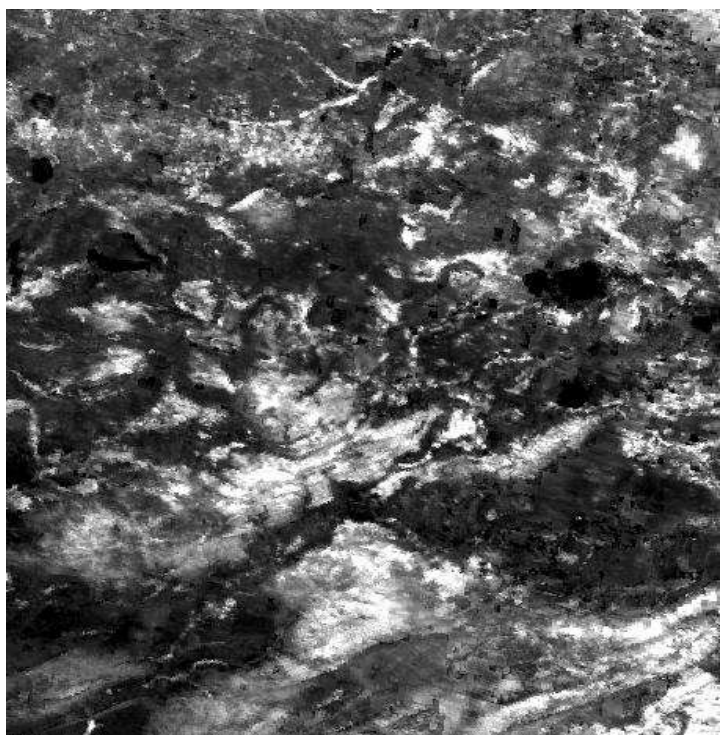

Fig. 8. NDVI MODIS image covering part of the Batna city taken on December 19, 2002

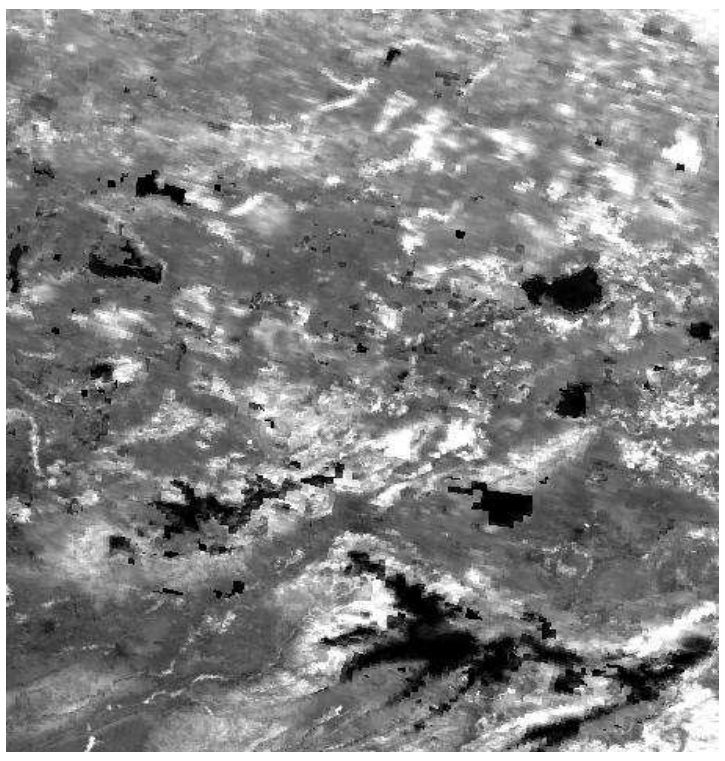

Fig. 9. NDVI MODIS image covering part of the Batna city taken on January 17, 2003

### *B. Results*

We present in this part the implementation steps of the application “SAX” on our base of SITS:

*1) Step 1: Data Storage*

This step consists to create a MySQL database in which all satellite MODIS NDVI images will be stored.

Images are identified principally by their acquisition date and size.

*2) Step2 : Images Time Series Representation*

This step is to represent the NDVI change in time of each pixel's time series , we obtained a total of 237055 time series of the same size (135 points). At each time $t_i$ representing the image acquisition date, NDVI value is affected.

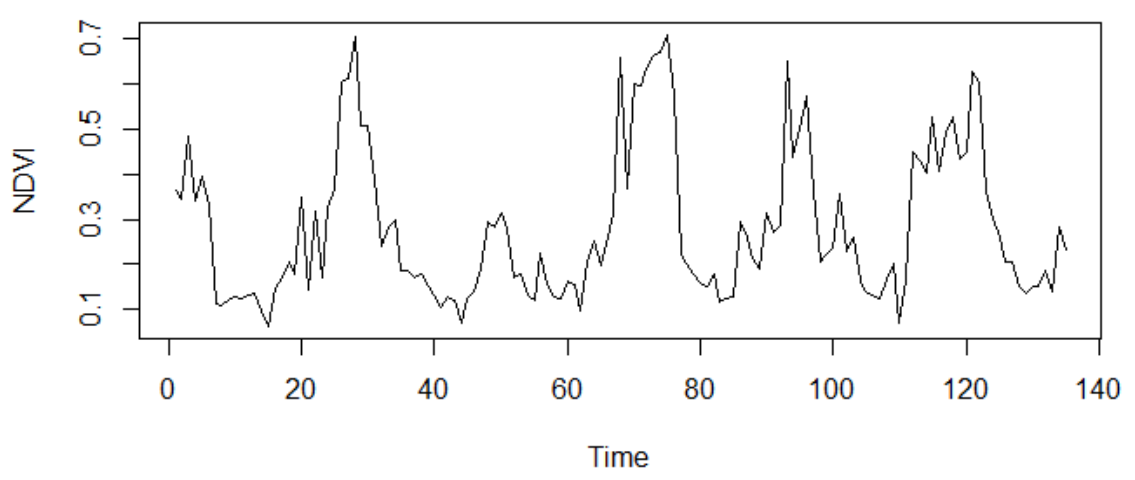

Fig. 10. Example of a time series of the first pixel NDVI evolution.

*3) Step3 : Symbolic Representation Time Series by SAX method*

In this step, each time series of length n, such that n = 135, is transformed into SAX symbolic sequence of length w as w << n.

The SAX parameters choice was defined as mentioned below:

The alphabet used is a finite set of natural numbers {1, 2, 3, 4, 5,.. , 20 ...}.

Alphabet size named Code Book (CB) will be selected by user, for example from the even numbers: 2,4,6….to 64, such as CB ⩾ 2. For example, for the alphabet size CB=4, symbols are selected from the set = {1, 2, 3, 4}. The Word Code (CW) size for which the time series will be divided into the same size segments will be selected by user from even numbers: 2, 4, 6… 32.

Then we combine the two parameter values, this gives the resolution (CB x CW): 2x2, 2x4, 2X6, 2X8 .......... 64X28, 64X30, 64x32. For each CBXCW resolution, the steps of the SAX method are applied to all SITS and we get in results SAX sequences.

First, the division of all SITS (obtained in step 2) into sub-sequences of equal length, for each time series C = $c_1$... $c_n$ (which $c_i$ represents NDVI), a sliding window CW traverses all points of the series and the division into the same CW intervals size. In each position ((i*1)*CW+1) (i*CW)), the average values of the points located in the current interval is calculated. And we get in result PAA vector $\bar{C} = \bar{c}_1, \ldots, \bar{c}_w$.

The second step consists to quantifying the PAA vectors by a symbol of our alphabet. This transformation is performed by assigning a symbol with equal probability to each element of PAA vector by referring to a Gaussian distribution. This is the assumption of the SAX method which suggests that the values of the time series are in N (0,1).

The Gaussian quantiles are calculated, for an alphabet size = N, we obtain (N-1) quantile values. Then we compare each value of the PAA vector with the Gaussian quantiles. For all the PAA vector values that are below the smallest quantile, the assigned symbol will be "1", for all the PAA vector values that they are greater than or equal to the smallest quantile and they are less small than the second smallest quantile; the assigned symbol will be "2"... etc.

Examples of Results:

CBXCW = 10X4

Pixel position "1" :

SAX sequence is 6 5 5 5 5 5 7 6 5 5 5 5 6 5 5 5 6 7 7 5 5 5 6 7 6 5 5 5 6 6 6 5 5

Pixel position " 203743":

SAX sequence is 7 7 6 6 6 7 8 7 6 6 6 8 7 6 5 6 6 7 7 5 5 5 6 7 7 6 5 6 7 8 6 5 5

Pixel position "237055":

SAX sequence is 7 7 7 8 8 8 7 7 8 8 8 8 7 8 8 8 8 7 6 7 8 8 8 7 7 7 8 8 8 7 7 8 8

CBXCW = 10X2

Pixel position "1":

SAX sequence is 6 6 6 5 5 5 5 5 5 6 5 5 6 7 6 6 6 5 5 5 5 5 5 5 6 5 5 5 5 5 5 5 6 6 7 7 7 5 5 5 5 5 5 5 6 7 7 6 5 6 5 5 5 5 6 6 6 6 6 7 6 5 5 5 5 5

Pixel position "203745":

SAX sequence is 7 8 7 6 5 5 5 5 5 6 6 7 8 7 8 6 5 5 5 5 5 5 6 7 8 7 6 5 5 5 5 5 5 6 7 7 8 7 5 5 5 5 5 5 6 6 7 8 8 6 5 5 5 5 6 6 7 7 7 7 7 5 5 5 5 5 5

Pixel position "237055":

SAX sequence is 7 6 7 7 7 8 8 8 8 8 8 7 7 7 7 8 8 8 7 8 8 8 7 7 7 8 8 8 8 8 8 8 8 8 7 6 7 7 8 8 8 8 8 8 8 7 6 6 7 7 8 8 8 8 8 8 8 7 7 7 7 8 8 8 8 8

CBXCW =32X10

Pixel position "1":

SAX sequence is 19 18 21 18 18 18 19 21 18 20 18 21 19

Pixel position "203745":

SAX sequence is 23 18 24 19 22 19 20 23 18 23 18 24 18

Pixel position "237055":

SAX sequence is 23 26 24 25 25 26 26 23 26 23 25 25 25

## VII. Conclusion

In order to better exploitation an important volume of satellite data, we applied the SAX method on satellite images time series (SITS) MODIS based on the vegetation index (NDVI) with an acquisition frequency of 16 days in a period of 5 years. First, SAX proceeds to transforming the SITS to PAA representation, then to symbolizing PAA representation to a discrete representation using numbers alphabet (1, 2, 3,..), and in function of CB and CW parameters' choice, several tests were performed and results are satisfactory and show that SAX is effective to reducing the dimensionality of SITS size and also fast even in the treatment of long time series

## References

[1] T. Guyet, H. Nicolas, and A. Diouck, "Segmentation multi-échelle de séries temporelles d'images satellite : Application à l'étude d'une période de sécheresse au Sénégal", RFIA 2012 (Reconnaissance des Formes et Intelligence Artificielle), pp.978-2-9539515-2-3, January 2012.

[2] B. Hugueney, "Représentations symboliques adaptatives de séries temporelles : principes et algorithmes de construction", Proceedings of the 6th EGC'2006 (Extraction et Gestion des Connaissances), pp.76-85, January 2006.

[3] S. Malinowski, “Enrichissement de la représentation SAX pour la fouille de séries temporelles“, Data Mining Day, March 2013.

[4] M. Marwan, M.Fuad, P. Marteau, “Towards A Faster Symbolic Aggregate Approximation Method”, ICSOFT 2010, Fifth International Conference on Software and Data Technologies, pp.305-310, July 2010.

[5] J.Lin, E.Keogh, L.Wei and S. Lonardi, “SAX: a Novel Symbolic Representation of Time Series”, Data Mining and Knowledge Discovery, Volume 15, pp.107-144, October 2007.

[6] E. Keogh and J. Lin, “Symbolic Representations of Time Series”, University of California, 2003.

[7] S. Malinowski, T.Guyet, R.Quiniou and R.Tavenard, “1d-SAX: a Novel Symbolic Representation for Time Series”, International Symposium on Intelligent Data Analysis, pp.273-284, 2013.

[8] J. Lin, E. Keogh, S.Lonardi and B. Chiu, “A Symbolic Representation of Time Series, with Implications for Streaming Algorithms”, DMKD '03 Proceedings of the 8th ACM SIGMOD workshop on Research issues in data mining and knowledge discovery, pp.2-11, 2003.